\documentstyle[12pt,epsf]{article}
\renewcommand{\bar}[1]{\overline{#1}}
\newcommand{\ie}{{\it i.e.}}

\begin{document}
 
\begin{flushright}
SLAC-PUB-7126\\
BIHEP-TH-96-13
\end{flushright}
 
\bigskip\bigskip
\centerline{{\Large \bf The Quark/Antiquark Asymmetry}}
\smallskip
\centerline{{\Large \bf  of the Nucleon Sea}
\footnote{\baselineskip=13pt
Work partially supported by the Department of Energy, contract
DE--AC03--76SF00515 and by National Natural Science Foundation of
China under Grant No.~19445004
and National Education Committee of China under Grant
No.~JWSL[1995]806.
}}
\vspace{42pt}
 
\centerline{\bf  Stanley J.~Brodsky}
\vspace{8pt}
\centerline{Stanford Linear Accelerator Center}
\centerline{Stanford University, Stanford, California 94309, USA}
\centerline{e-mail: sjbth@slac.stanford.edu}
\vspace*{8pt}
\centerline{\bf Bo-Qiang Ma}
\centerline{CCAST (World Laboratory), P.O.~Box 8730, Beijing 100080,
China}
\centerline{Institute of High Energy Physics, Academia Sinica}
\centerline{P.~O.~Box 918(4), Beijing 100039, China}
\centerline{and}
\centerline{Faculty of Natural Sciences, Kwanghua Academy of
Sciences}
\centerline{Beijing 100081, China}
\centerline{e-mail: mabq@bepc3.ihep.ac.cn}
\vfill
\centerline{Submitted to Physics~Letters~B.}
\vfill
\newpage
 
\vspace{40pt}
\begin{center}
{\Large \bf Abstract}
\end{center}
 
Although the distributions of sea quarks and antiquarks generated by
leading-twist QCD evolution through gluon splitting $g \rightarrow
\bar q q$ are necessarily CP symmetric, the distributions of
nonvalence quarks and antiquarks which are intrinsic to the
nucleon's bound state wavefunction need not be identical. In this
paper we investigate the sea quark/antiquark asymmetries in the
nucleon wavefunction which are generated by a light-cone model of
energetically-favored meson-baryon fluctuations.  The model predicts
striking quark/antiquark asymmetries in the momentum and helicity
distributions for the down and strange contributions to the proton
structure function: the intrinsic $d$ and $s$ quarks in the proton
sea are predicted to be negatively polarized, whereas the intrinsic
$\bar d$ and $\bar s$ antiquarks give zero contributions to the
proton spin. Such a picture is supported by experimental phenomena
related to the proton spin problem  and the violation of  the
Ellis-Jaffe sum rule. The light-cone meson-baryon fluctuation model
also suggests a structured momentum distribution asymmetry for
strange quarks and antiquarks which could be relevant to an
outstanding  conflict between two different determinations of the
strange quark sea in the nucleon. The model predicts an excess of
intrinsic $d \bar d$ pairs over $u \bar u$ pairs, as supported by
the Gottfried sum rule violation. We also predict that the intrinsic
charm and anticharm helicity and momentum distributions are not
identical.
 
\vfill
\centerline{PACS numbers: 13.60.Hb, 11.30.Hv, 12.39.Ki, 13.88.+e}
\vfill
\newpage
 
\section{Introduction}
 
The composition of the nucleon in terms of its fundamental quark and
gluon degrees of freedom is a key focus of QCD phenomenology. In the
light-cone (LC) Fock state description of bound states, each
hadronic eigenstate of the QCD Hamiltonian is expanded at fixed
light-cone time $\tau = t + z/c$ on the complete set of
color-singlet eigenstates $\{ \vert n  \rangle \}$ of the free
Hamiltonian which have the same global quantum numbers:   $\vert  p
\rangle = \Sigma_n \psi_n^H(x_i, k_{\perp i}, \lambda_i) \vert n
\rangle.$  Thus each Fock component in the light-cone wavefunction
of a nucleon  is composed of three valence quarks, which gives the
nucleon its global quantum numbers, plus a variable number of sea
quark-antiquark ($q \bar q$) pairs of any flavor, plus any number of
gluons.   The quark distributions $q(x,\widetilde Q)$ of the nucleon
measured in deep inelastic scattering are computed from the sum of 
squares of the light-cone wavefunctions integrated over transverse
momentum $k_\perp$ up to the factorization scale $\widetilde Q$,
where the light-cone momentum fraction $x = {k^+ \over p^+} =
{k^0+k^3\over P^0 + P^3}$ of the struck quark  is set equal to the
Bjorken variable $x_{BJ}.$
 
It is important to distinguish two distinct types of quark and gluon
contributions to the nucleon sea measured in deep inelastic
lepton-nucleon scattering: ``extrinsic" and ``intrinsic"
\cite{Bro81,Bur92,Bro95}. The extrinsic sea quarks and gluons are
created as part of the lepton-scattering interaction and thus exist
over a very short time $\Delta \tau \sim 1/Q$. These factorizable
contributions can be systematically derived from the QCD hard
bremsstrahlung and pair-production (gluon-splitting) subprocesses
characteristic of leading twist perturbative QCD evolution. In
contrast, the intrinsic sea quarks and gluons are multiconnected to
the valence quarks and exist over a relatively long lifetime within
the nucleon bound state. Thus the intrinsic $q \bar q$ pairs can
arrange themselves together with the valence quarks of the target
nucleon into the most energetically-favored meson-baryon
fluctuations.
 
In conventional studies of the ``sea'' quark distributions, it is
usually assumed that, aside from the effects due to
antisymmetrization, the quark and antiquark sea contributions have
the same momentum and helicity distributions. However, the ansatz of
identical quark and antiquark sea contributions has never been
justified, either theoretically or empirically. Obviously the sea
distributions which arise directly from gluon splitting in leading
twist are necessarily CP-invariant; \ie,\ they are symmetric under
quark and antiquark interchange. However, the initial distributions
which provide the boundary conditions for QCD evolution need not be
symmetric since the nucleon state is itself  not CP-invariant. Only
the global quantum numbers of the nucleon must be conserved. The
intrinsic sources of strange (and charm) quarks reflect the
wavefunction structure of the bound state itself; accordingly, such
distributions would not be expected to be CP symmetric. Thus the
strange/antistrange asymmetry of nucleon structure  functions
provides a direct window into the  quantum bound-state structure of
hadronic wavefunctions.
 
It is also possible to consider the nucleon wavefunction at low
resolution as a fluctuating system coupling  to intermediate
hadronic Fock states such as noninteracting meson-baryon pairs. The
most important fluctuations are most likely to be those closest to
the energy shell and thus have minimal invariant mass.  For example,
the coupling of a proton to a virtual $K^+ \Lambda$ pair provides a
specific source of intrinsic strange quarks and antiquarks in the
proton.   Since the $s$ and $\bar s$ quarks appear in different
configurations in the lowest-lying hadronic pair states, their
helicity and momentum distributions are distinct.
 
The purpose of this paper is to investigate the quark and antiquark
asymmetry in the nucleon sea which is implied by a light-cone
meson-baryon fluctuation model of intrinsic $q\bar q$ pairs. Such
fluctuations are necessarily part of any  quantum-mechanical
description of the hadronic bound state in QCD and have also been
incorporated into the cloudy bag model \cite{Sig87} and Skyrme
solutions to chiral theories \cite{Bur92}. We shall utilize a
boost-invariant light-cone Fock state description of the hadron
wavefunction which emphasizes multi-parton configurations  of
minimal invariant mass. We find that such fluctuations predict a
striking sea quark and antiquark asymmetry in the corresponding
momentum and helicity distributions in the nucleon structure
functions.  In particular, the strange and antistrange distributions
in the nucleon generally have completely different momentum and spin
characteristics. For example, the model predicts that the intrinsic
$d$ and $s$ quarks in the proton sea are negatively polarized,
whereas the intrinsic $\bar d$ and $\bar s$ antiquarks provide zero
contributions to the proton spin. We also predict that the intrinsic
charm and anticharm helicity and momentum distributions are not
strictly identical. We show that the above picture of quark and
antiquark asymmetry in the momentum and helicity distributions of
the nucleon sea quarks has support from a number of experimental
observations, and we suggest processes to test and measure this
quark and antiquark asymmetry in the nucleon sea.
 
\section{The light-cone meson-baryon fluctuation model of intrinsic
$q \bar q$ pairs}
 
In order to characterize the momentum and helicity distributions of
intrinsic $q \bar q$ pairs, we shall adopt a light-cone two-level
convolution model of structure functions \cite{Ma91} in which the
nucleon is a two-body system of meson and baryon which are also
composite systems of quarks and gluons. We first study the intrinsic
strange $s \bar s$ pairs in the proton.  In the meson-baryon
fluctuation model, the intrinsic strangeness fluctuations in the
proton wavefunction are mainly due to the intermediate $K^+ \Lambda$
configuration since this state has the lowest off-shell light-cone
energy and invariant mass \cite{Bro95}. The  $K^+$ meson is a
pseudoscalar particle with negative parity, and the $\Lambda$ baryon
has the same parity of the nucleon. The $K^+ \Lambda$ state retains
the  parity of the proton,  and  consequently,  the $K^+ \Lambda$
system must have odd orbital angular momentum.  We thus write the
total angular momentum space wavefunction of the intermediate $K
\Lambda$ state in the center-of-mass reference frame as
\begin{eqnarray}
\left|J=\frac{1}{2},J_z=\frac{1}{2}\right\rangle
&=&\sqrt{\frac{2}{3}} \left|L=1,L_z=1\right\rangle\
\left|S=\frac{1}{2},S_z=-\frac{1}{2}\right\rangle\nonumber\\
&-&\sqrt{\frac{1}{3}}\
\left|L=1,L_z=0\right\rangle \
\left|S=\frac{1}{2},S_z=\frac{1}{2}\right\rangle\ .
\end{eqnarray}
In the constituent quark model,  the spin of $\Lambda$ is provided
by its strange quark and the net spin of the antistrange quark in
$K^+$ is zero. The net spin projection of $\Lambda$ in the
$K^+\Lambda$ state is $S_z(\Lambda)=-\frac{1}{6}$. Thus the
intrinsic strange quark normalized to the probability
$P_{K^+\Lambda}$ of the $K^+\Lambda$ configuration yields a
fractional contribution $\Delta S_{s}=2
S_z(\Lambda)=-\frac{1}{3}P_{K^+\Lambda} $ to the proton spin,
whereas the intrinsic antistrange quark gives a zero contribution:
$\Delta S_{\bar s}=0.$ There thus can be a significant quark and
antiquark asymmetry in the quark spin distributions for the
intrinsic $s \bar s$ pairs.
 
We also need to estimate the relative probabilities for other
possible fluctuations with higher off-shell light-cone energy and
invariant mass, such as $K^+(u\bar{s}) \Sigma^0(uds)$,
$K^0(d\bar{s})\Sigma^+(uus)$, and $K^{* +}(u\bar s) \Lambda(uds)$.
For example, we find that the relative probability of finding the
$K^{* +} \Lambda$ configuration (which has positively-correlated
strange quark spin) compared with $K^+ \Lambda$ is 3.6\% (8.9\%) by
using a same normalization constant for a light-cone Gaussian type
(power-law type) wavefunction. We find that the higher fluctuations
of intrinsic $s \bar s$ pairs do not alter the qualitative estimates
of the quark and antiquark spin asymmetry in the nucleon sea based
on using the $K^+ \Lambda$ fluctuation alone.
 
The quark helicity projections measured in deep inelastic scattering
are related to the quark spin projections in the target rest frame
by multiplying by a Wigner rotation factor of order 0.75 for light
quarks and of order 1 for heavy quarks \cite{Ma91b}.  We therefore
predict that the net strange quark helicity arising from the
intrinsic $s \bar s$ pairs in the nucleon wavefunction is negative,
whereas the net antistrange quark helicity is approximately zero.
This aspect of quark/antiquark helicity asymmetry is in qualitative
agreement with the predictions of a broken-U(3) version of the
chiral quark model \cite{Man84} where the intrinsic quark-antiquark
pairs are introduced through quark to quark and Goldstone boson
fluctuations.
 
In principle, one can measure the helicity distributions of strange
quarks in the nucleon sea from the $\Lambda$ longitudinal
polarization of semi-inclusive $\Lambda$ production in polarized
deep inelastic scattering \cite{Bro95,Lu95,Ell95}.  We expect that
the polarization is negative for the produced $\Lambda$ and zero (or
slightly positive \cite{Bur92}) for the produced $\bar{\Lambda}$.
The expectation of negative longitudinal $\Lambda$ polarization is
supported by the measurements of the WA59 Collaboration for the
reaction $\bar{\nu}+{\rm N} \rightarrow\mu^+\Lambda+{\rm X}$
\cite{WA59}. The complementary measurement of $\bar {\Lambda}$
polarization in semi-inclusive deep inelastic scattering is clearly
important in order to test the physical picture of the light-cone
meson-baryon fluctuation model.
 
It is also interesting to study the sign of the $\Lambda$
polarization in the current fragmentation region as the Bjorken
variable $x \rightarrow 1$ in polarized proton deep inelastic
inclusive reactions. From perturbative QCD arguments on helicity
retention, one expects a positive helicity distribution for any
quark struck in the end-point region $x \rightarrow 1$, even though
the global helicity correlation is negative \cite{Bro95b}. Thus we
predict that the sign of the $\Lambda$ polarization should  change
from negative to positive as $x$ approaches the endpoint regime.
 
\begin{figure}[htbp] 
\begin{center} 
\leavevmode
{\epsfxsize=4in \epsfbox{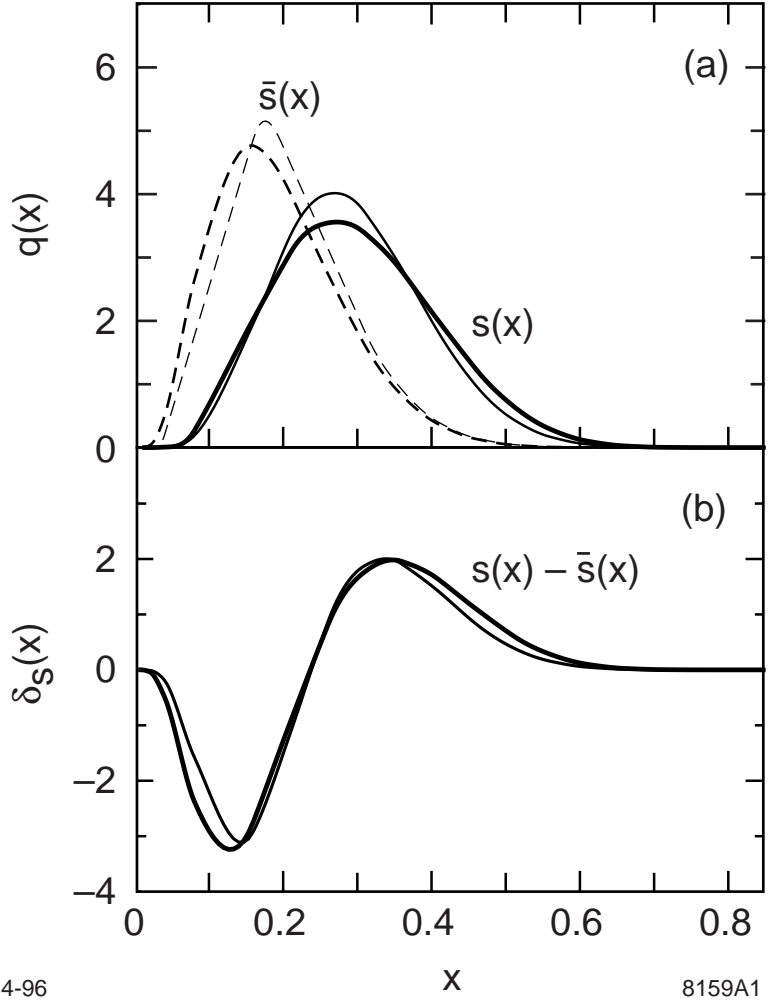}} 
\end{center}
\caption{\baselineskip 13pt The momentum distributions for the
strange quarks and antiquarks in the light-cone meson-baryon
fluctuation model of intrinsic $q \bar q$ pairs, with the
fluctuation wavefunction of $K^+\Lambda$ normalized to 1. The curves
in (a) are the calculated results of $s(x)$ (solid curves) and $\bar
s(x)$ (broken curves) with the Gaussian type (thick curves) and
power-law type (thin curves) wavefunctions and the curves in (b) are
the corresponding $\delta_s(x)=s(x)-\bar s(x)$. The parameters are
$m_q=330$ MeV for the light-flavor quark mass, $m_s=480$ MeV for the
strange quark mass, $m_D=600 $ MeV for the spectator mass, the
universal momentum scale $\alpha=330$ MeV, and the power constant
$p=3.5$, with realistic meson and baryon masses. } 
\label{bmfig1}
\end{figure}

The momentum distributions of the intrinsic strange and antistrange
quarks in the $K^+ \Lambda$ state can be modeled from the two-level
convolution formula
\begin{equation}
s(x)=\int_{x}^{1} \frac{{\rm d}y}{y}
f_{\Lambda/K^+\Lambda}(y) q_{s/\Lambda}\left(\frac{x}{y}\right);
\;\;\;
\bar s(x)=\int_{x}^{1} \frac{{\rm d}y}{y}
f_{K^+/K^+\Lambda}(y) q_{\bar s/K^+}\left(\frac{x}{y}\right),
\end{equation}
where $f_{\Lambda/K^+\Lambda}(y)$, $f_{K^+/K^+\Lambda}(y)$ are
probabilities of finding $\Lambda$, $K^+$ in the $K^+ \Lambda$ state
with the light-cone momentum fraction $y$ and $q_{s/\Lambda}(x/y)$,
$q_{\bar s/K^+}(x/y)$ are probabilities of finding strange,
antistrange quarks in $\Lambda$, $K^+$ with the light-cone momentum
fraction $x/y$. We shall estimate these quantities by adopting
two-body momentum wavefunctions for $p=K^+ \Lambda$, $K^+=u {\bar
s}$, and $\Lambda=s u d$ where the $u d$ in $\Lambda$ serves as a
spectator in the quark-spectator model \cite{Ma96}. We choose two
simple functions of the invariant mass $ {\cal M}^2=\sum_{i=1}^{2} \
\frac{{\bf k}^2_{\perp i}+m_i^2}{x_i} $ for the two-body
wavefunction: the Gaussian type and power-law type wavefunctions
\cite{BHL}, \begin{equation} \psi_{{\rm Gaussian}}({\cal
M}^2)=A_{{\rm Gaussian}}\  {\rm exp} (-{\cal M}^2/2\alpha^2),
\end{equation} \begin{equation} \psi_{{\rm Power}}({\cal
M}^2)=A_{{\rm Power}}\ (1+{\cal M}^2/\alpha^2)^{-p}, \end{equation}
where $\alpha$ sets the characteristic internal momentum scale. We
do not expect to produce realistic quark distributions with simple
two-body wavefunctions;  however, we can hope to explain some
qualitative features of the data as well as relations between the
different quark distributions \cite{Ma96}. The predictions for the
momentum distributions $s(x)$, $\bar s(x)$, and $\delta
_s(x)=s(x)-\bar s(x)$ are presented in Fig.~\ref{bmfig1}.    The
strong, structured momentum asymmetry of the $s(x)$ and $\bar s(x)$
intrinsic distributions reflects the tendency of the wavefunction 
to minimize the relative velocities of the intermediate meson baryon
state and is approximately  same for the two wavefunctions. 
\vspace{.5cm} 
 
\begin{figure}[htbp] 
\begin{center}
\leavevmode 
{\epsfxsize=4in\epsfbox{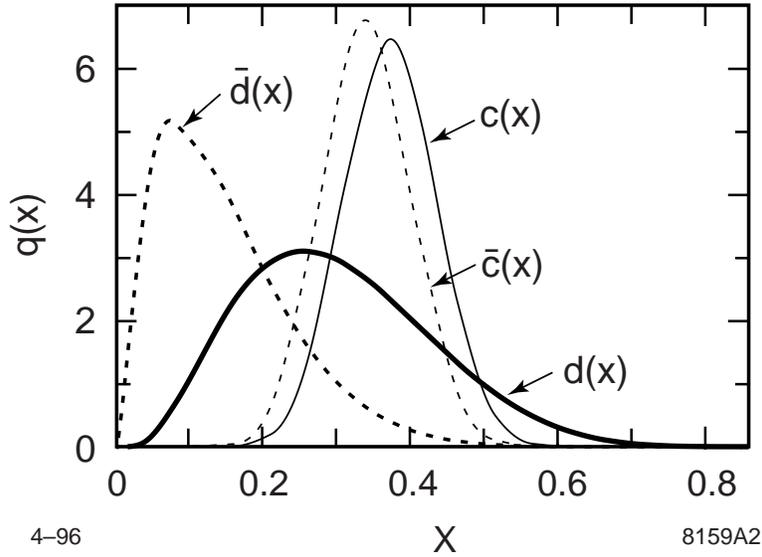}} 
\end{center}
\caption{\baselineskip 13pt The momentum distributions for the down
and charm quarks and antiquarks in the light-cone meson-baryon
fluctuation model of intrinsic $q \bar q$ pairs, with the
fluctuation wavefunctions normalized to 1. The curves are the
calculated results for $d(x)$ (thick solid curve), $\bar d(x)$
(thick broken curve), $c(x)$ (thin solid curve), and $\bar c(x)$
(thin broken curve) with the Gaussian type wavefunction. The
parameters are $m_d=330$ MeV for the down quark mass and $m_c=1500$
MeV for the charm quark mass, with other parameters as those in
Fig.~1. } 
\label{bmfig2} 
\end{figure}

We have performed similar calculations for  the momentum
distributions of the intrinsic $d \bar d$ and $c \bar c$ pairs
arising from the $p(uud)=\pi^{+}(u\bar{d})n(udd)$ and
$p(uud)=\bar{D}^0(u\bar c) \Lambda^{+}_{c}(udc)$ configurations. The
results are presented in Fig.~\ref{bmfig2}. We find a large quark
and antiquark asymmetry for the $d \bar d$ pairs. The $c \bar{c}$
momentum asymmetry is small compared with the $s \bar{s}$ and $d
\bar{d}$ asymmetries but is still nontrivial. The $c \bar{c}$ spin
asymmetry, however, is large. Considering that it is difficult to
observe the momentum asymmetry for the $d \bar d$ pairs due to an
additional valence $d$ quark in the proton, the momentum asymmetry
of the intrinsic strange and antistrange quarks is the  most
significant feature of the model and the easiest to observe.  We
have not taken into account the antisymmetrization effect of the $u$
and $d$ sea quarks with the valence quarks \cite{Bro91}.

\section{The strange quark/antiquark asymmetry in the nucleon
sea}
 
We have seen that the light-cone meson-baryon fluctuation model of
intrinsic $q \bar q$ pairs leads to significant quark/antiquark
asymmetries in the momentum and helicity distributions of the
nucleon sea quarks. A strange/antistrange asymmetry in the nucleon
sea has also been suggested from estimates in the cloudy bag model
\cite{Sig87} and Skyrme solutions to chiral theories \cite{Bur92}.
At present there is still no direct experimental confirmation of the
strange/antistrange asymmetry. However, a strange/antistrange
momentum asymmetry in the nucleon can be inferred from the apparent
conflict between two different determinations of the strange quark
content in the nucleon sea.
 
\begin{figure}[htbp]
\begin{center}
\leavevmode
{\epsfxsize=4in\epsfbox{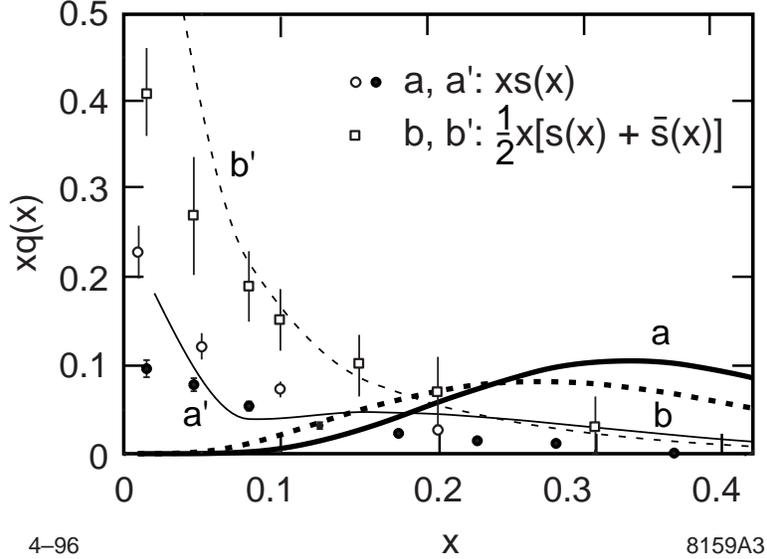}}
\end{center}
\caption[*]{\baselineskip 13pt
Results for the strange quark distributions $x s(x)$ and $x \bar
s(x)$ as a function of the Bjorken scaling variable $x$. The open
squares shows the CTEQ determination of $\frac{1}{2}\,x[s(x)+\bar
s(x)]$ obtained from $\frac{5}{12}\,(F_{2}^{\nu{\cal
N}}+F_{2}^{\overline{\nu}{\cal N}})(x)({\rm CCFR})- 3F_{2}^{\mu
{\cal N}}(x)({\rm NMC}).$  The circles show the CCFR determinations
for $x s(x)$ from dimuon events in neutrino scattering using a 
leading-order QCD analysis at $Q^{2} \approx 5 ({\rm GeV/c})^{2} $
(closed circles) and a  higher-order QCD analysis at $Q ^2 =20 ({\rm
GeV/c})^2$ (open circles). The thick curves are the unevolved
predictions of the light-cone fluctuation model for $x s(x)$ (solid
curve labeled a) and $\frac{1}{2}x[s(x)+\bar s(x)]$ (broken curve
labeled b) for the Gaussian type wavefunction in the light-cone
meson-baryon fluctuation model of intrinsic $q \bar q$ pairs
assuming a  probability of 10\% for the $K^+ \Lambda$ state. The
thin solid and broken curves (labeled a' and b') are the
corresponding evolved predictions multiplied by the factor
$d_v(x)|_{fit}/d_v(x)|_{model}$ assuming  a probability of 4\% for
the $K^+ \Lambda$ state. } 
\label{bmfig3} 
\end{figure}

The strange quark distribution in the nucleon is usually obtained
from  analyses of the deep inelastic lepton-nucleon scattering data
assuming identical momentum distribution for the strange and
antistrange quark distributions, \ie, $s(x)=\bar s(x)$. The CTEQ 
\cite{CTEQ93} global analyses of quark distributions are based
primarily  on the following representations of the nucleon structure
functions: 
\begin{eqnarray} 
F_{2}^{\mu p}-F_{2}^{\mu n}&=&\frac{1}{3}\,x(u+\overline{u}-d-
\overline{d}); \\ F_{2}^{\mu {\cal N} }=\frac{1}{2}\,(F_{2}^{\mu
p}+F_{2}^{\mu n})&=&\frac{5}{18}\,x
\left[u+\overline{u}+d+\overline{d}+\frac{2}{5}\,(s+\bar s)\right];
\label{eq:eq2} \\ 
\frac{1}{2}\,(F_{2}^{\nu {\cal N}}+F_{2}^{\overline{\nu} {\cal N}})
&=&x(u+\overline{u}+d+\overline{d}+s+\bar s); \label{eq:eq3} \\
\frac{1}{2}\,(F_{3}^{\nu {\cal N}}+F_{3}^{\overline{\nu} {\cal
N}})&=& (u-\overline{u}+d-\overline{d}+s-\overline{s}),
\label{eq:eq4} 
\end{eqnarray} 
where $F_{2,3}^{{\cal N}}$ are converted from  $F_{2,3}^{Fe}$ using
a heavy-target correction factor, with $F_{2,3}^{{\cal N}}$ denoting
$\frac{1}{2}(F_{2,3}^{p} +F_{2,3}^{n})$. These four observables
determine four combinations of quark distributions, which can be
taken to be $u+\overline{u}$, $d+\overline{d}$,
$\overline{u}+\overline{d}$ and $s+\bar s$ if one neglects the
relatively small contribution from the $s-\overline{s}$ term in
Eq.~(\ref{eq:eq4}). From Eqs.~(\ref{eq:eq2}) and (\ref{eq:eq3}), one
obtains the equality 
\begin{equation} 
\frac{5}{12}\,(F_{2}^{\nu{\cal N}}+F_{2}^{\overline{\nu}{\cal N}})-
3F_{2}^{\mu {\cal N}}=\frac{1}{2}\,x[s(x)+\bar s(x)]. 
\label{eq:eq5}
\end{equation} 
The CTEQ Collaboration has determined the quantity on the left-hand
side of Eq.~(\ref{eq:eq5}) at $Q^{2}=5$ GeV$^{2}$ using data from
the New Muon Collaboration (NMC) \cite{NMC92} and CCFR
\cite{CCFR92}, as shown in Fig.~\ref{bmfig3}.  The  strange quark
distribution measured in this way  can be  effectively identified
with the mean $\frac{1}{2}[s(x)+\bar s(x)]$.
 
\vspace{.5cm}
 
In principle, the difference between $s(x)$ and $\bar s(x)$ can be
determined from measurements of deep inelastic scattering by
neutrino and antineutrino beams since they probe strange and
antistrange quarks in distinct ways.  One also requires explicit
light-flavor quark distributions $u(x)$, $d(x)$ and $\bar u(x)$,
$\bar d(x).$   The CCFR determinations of the strange quark
distributions are obtained from 5044 neutrino and 1062 antineutrino
dimuon events by assuming $s(x)=\bar s(x)$ \cite{CCFR93}. However,
we can regard their results as a rough estimate of $s(x)$ alone
since the neutrino events dominate the data set.
 
In Fig.~\ref{bmfig3} we plot the calculated $x s(x)$ and
$\frac{1}{2}x[s(x)+\bar s(x)]$ where the normalization of the $K^+
\Lambda$ probability is adjusted to fit the measured strange sea at
$x \approx 0.2$. This prediction does not take into account QCD
evolution but indicates the inherent quark/antiquark asymmetry
predicted by the fluctuation model with the input light-cone
wavefunction. In order to reflect the evolution effect we have
simply multiplied by a factor $d_v(x)|_{fit}/d_v(x)|_{model}$ in the
light-cone quark-spectator model \cite{Ma96} to the calculated
$s(x)$ and $\bar s(x)$ in our model. Aside from very small $x$ where
shadowing may be playing a role, the predictions of the
model are in reasonable qualitative agreement with the empirical
determinations of $x s(x)$ and $\frac{1}{2}x[s(x)+\bar s(x)].$  Thus
the quark/antiquark asymmetry of the intrinsic strange quark
distributions may help to explain the apparent conflict between the
two measures. Of course, the actual wavefunctions underlying the
higher Fock states are undoubtedly more complicated than the simple
forms used here.
 
\begin{figure}[htbp]
\begin{center}
\leavevmode
{\epsfxsize=4in\epsfbox{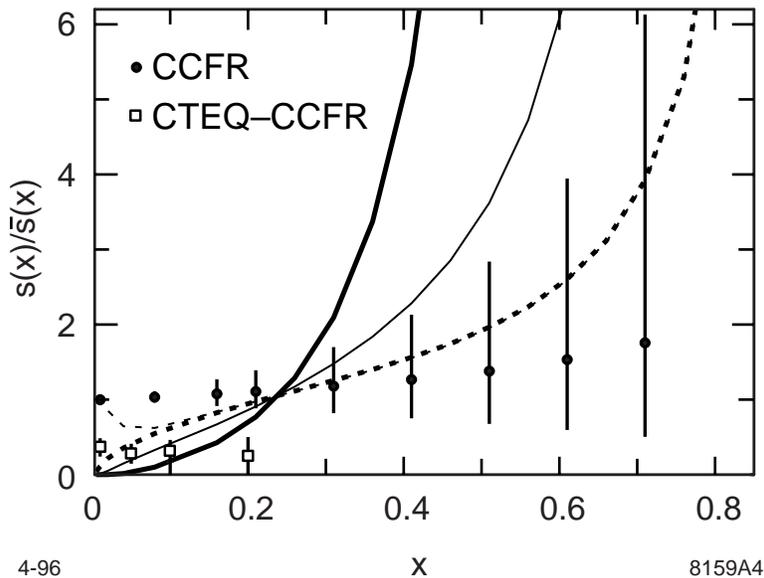}}
\end{center}
\caption{\baselineskip 13pt
Results for the strange asymmetry $s(x)/\bar s(x)$ as a function
of the Bjorken scaling variable $x$. The open squares are the
combined CTEQ-CCFR ``data'' and the closed circles are the CCFR
measurement of $s(x)/\bar s(x)$. The thick (thin) solid curve is the
calculated result of $s(x)/\bar s(x)$ in the light-cone meson-baryon
fluctuation model for the Gaussian type wavefunction with
$\alpha=330$ ($530$) MeV. The thick broken curve is the result with
a larger $\alpha=800$ MeV and the thin broken curve is the above
result with of about 30\% extrinsic strange quarks (\ie, $x
s_{extrinsic}=0.07(1-x)^5$) included for comparison with the CCFR
result at small $x$. } 
\label{bmfig4} 
\end{figure}

The CCFR collaboration has re-analyzed the strange quark
distributions within the context of a higher-order QCD analysis
\cite{CCFR95}. The discrepancy between the new measured strange
quark distributions and those of the CTEQ parametrizations is
reduced, but it is still significant. The CTEQ ``data" is still
larger than the new CCFR data,  in agreement with the ratio of 
$\frac{1}{2} x[s(x)+\bar s(x)]$ to $x s(x)$ predicted by the 
fluctuation model. In Fig.~\ref{bmfig4} we plot the CTEQ-CCFR
``data" for the strange asymmetry $s(x)/\bar s(x)$ by combining the
CTEQ ``data" and the new CCFR data. Allowing for the experimental
errors, there is a reasonable agreement between the ``data" and the
predictions of the LC fluctuation model if we increase the value for
the wavefunction momentum scale $\alpha$. A larger momentum scale
$\alpha$ is also required for a good description of structure
function within the light-cone quark model \cite{Hua94}. We should
be cautious about comparisons in the small $x $ region, since the
dominant contribution to strange and antistrange quarks in this
region comes from the extrinsic sea quarks as well as the evolution
of the intrinsic contributions; in fact, it has been shown that
gluon splitting alone is sufficient to describe the data \cite{Ji95}.
 
The CCFR collaboration has made a $s(x) \neq \bar s(x)$ fit to their
neutrino and antineutrino induced dimuon events and did not find
clear evidence for a difference between $s(x)$ and $\bar s(x)$
\cite{CCFR95}. However,  within the allowed errors, the CCFR data
for $s(x)/\bar s(x)$ does not rule out a strange asymmetry of the
magnitude suggested by the LC fluctuation model, see
Fig.~\ref{bmfig4}. It is also important to note that the CCFR
analysis of $s(x)$ and $\bar s(x)$ forces the data to fit specific
power-law parametrizations which preclude a structured asymmetric
intrinsic sea contribution of the type predicted by the LC
fluctuation model. Thus the CCFR parametrizations shown in Figs.~3
and 4 may not accurately represent the actual physics and we urge a
reanalysis that allows a structured asymmetric intrinsic strange
sea. Higher precision data and more studies are clearly needed in
order to positively confirm an asymmetry in the momentum
distributions of the strange and antistrange
 quarks.
 
\vspace{.5cm}
 
The normalization for intrinsic $s \bar s$ fluctuations can be
determined by fitting the two ``measurements" of the strange quark
distributions in the nucleon at $x \approx 0.2.$  We find a
probability of approximately 10\% for intrinsic $s \bar s$ pairs.
The net spin fraction of the intrinsic strange sea ($s_s$) quarks is
$\Delta S_{s_s}=-\frac{1}{3}$  normalized to the $s \bar s$
fluctuation of $K^+\Lambda$.   The  Wigner rotation factor should be
close to 1 due to the larger mass of the strange quarks. Thus the
helicity contribution from the intrinsic $s_s$ quarks with Fock
state probability 4-15\% is $\Delta s_s \approx -0.01$ to $0.05$,
which is somewhat smaller than the currently estimated empirical
value \cite{Ell95b} $\Delta s=-0.10 \pm 0.03$.  It is possible that
part of this discrepancy is due to the use of exact $SU(3)$ flavor
symmetry in the global fit \cite{Ell95b}, as suggested in
Refs.~\cite{SU3}.
 
\section{The light-flavor quark content in the nucleon sea}
 
The light-cone fluctuation model contains neutral meson fluctuation
configurations in which the intermediate mesons are composite
systems of the intrinsic up $u \bar u$ and down $d \bar d$ pairs,
but these fluctuations do not cause a quark/antiquark asymmetry in
the nucleon sea. The lowest non-neutral $u \bar u$ fluctuation in
the proton is $p(uud)=\pi^{-}(d \bar u)\Delta^{++}(uuu)$,  and its
probability is small compared to the non-neutral $d \bar d$
fluctuation. Therefore the dominant non-neutral light-flavor $q \bar
q$ fluctuation in the proton sea is $d \bar d$ through the
meson-baryon configuration $p(uud)=\pi^{+}(u \bar{d})n(udd)$. This
leads naturally to an excess of $d \bar d$ pairs over $u \bar u$
pairs in the proton sea.  Such a mechanism provides a natural
explanation \cite{Pi} for the violation of the Gottfried sum rule
\cite{NMC91} and leads to non-trivial distributions of the sea
quarks. The NMC measurement $ S_G=\frac{1}{3}+\frac{2}{3}
\int_0^1{\rm d} x [u_s(x)-d_s(x)] =0.235 \pm 0.026 $ \cite{NMC91}
implies $\int_0^1{\rm d} x [d_s(x)-u_s(x)]=0.148\pm 0.039$, which
can be considered as the probability of finding non-neutral
intrinsic $d \bar d$ fluctuations in the proton sea.
 
In the light-cone meson-baryon fluctuation model, the net $d$ quark
helicity of the intrinsic $q \bar q$ fluctuation is negative,
whereas the net $\bar d$ antiquark helicity is zero. Therefore the
quark/antiquark asymmetry of the $d \bar d$ pairs should be 
apparent in the $d$ quark and antiquark helicity distributions.
There are now explicit measurements of the helicity distributions
for the individual $u$ and $d$ valence and sea quarks by the Spin
Muon Collaboration (SMC) \cite{NSMCN}.  The helicity distributions
for the $u$ and $d$ antiquarks are consistent with zero in agreement
with the results of the light-cone meson-baryon fluctuation model of
intrinsic $q \bar q$ pairs.
 
The explicit $x$-dependent helicity distributions for valence $u$
and $d$ quarks can be related to the unpolarized valence quark
distributions in a light-cone SU(6) quark-spectator model
\cite{Ma96} for nucleon structure functions by taking into account
the flavor asymmetry due to the difference between scalar and vector
spectators and the Wigner rotation effect from the quark transversal
motions \cite{Ma91b}. Although the SMC data have fairly large
errors,  the calculated $\Delta u_v(x)$ in the light-cone
quark-spectator model are in good agreement with the data. However,
the predicted $\Delta d_v(x)$ distributions do not fit the data well
and seem to demand an additional negative contribution. This again
supports the light-cone meson-baryon fluctuation model in which the
helicity distribution of the intrinsic $d$ sea quarks $\Delta
d_s(x)$ is negative.
 
The standard SU(6) quark model gives the constraints $|\Delta u_v|
\leq \frac{4}{3}$ and $|\Delta d_v| \leq \frac{1}{3}$. A global fit
\cite{Ell95b} of polarized deep inelastic scattering data together
with constraints from nucleon and hyperon decay and the included
higher-order perturbative QCD corrections leads to values for
different quark helicity contributions in the proton: $\Delta
u=0.83\pm 0.03, \;\; \Delta d=-0.43 \pm 0.03, \;\; \Delta s=-0.10
\pm 0.03.$ In the light-cone meson-baryon fluctuation model, the
antiquark helicity contributions are zero. We thus  can consider the
empirical values  as the helicity contributions $\Delta q=\Delta
q_v+\Delta q_s$ from both the valence $q_v$ and sea $q_s$ quarks.
Thus the empirical result  $|\Delta d| > \frac{1}{3}$ strongly
implies an additional negative contribution $\Delta d_s$ in the
nucleon sea.
 
\section{Summary}
 
Intrinsic sea quarks clearly play a key role in determining basic
properties of the nucleon, including its static measures such as the
strange quark contribution to the nucleon magnetic moments.  As we
have shown here, the corresponding intrinsic contributions to the
sea quark structure functions lead to nontrivial, asymmetric, and 
structured momentum and spin distributions. Understanding the role
of the intrinsic distributions is essential for setting the boundary
conditions for the QCD evolution of the quark sea.  Although the sea
quarks generated by perturbative leading twist QCD evolution from
gluon spitting are necessarily CP and flavor symmetric, this is not
true for the momentum and helicity distributions of intrinsic sea
quarks controlled by the bound state nature of the hadrons.
 
In this paper we have studied the sea quark/antiquark asymmetries in
the nucleon wavefunction which are generated by a light-cone model
of energetically-favored meson-baryon fluctuations.  The model
predicts striking quark/antiquark asymmetries in the momentum and
helicity distributions for the down and strange contributions to the
proton structure function: the intrinsic $d$ and $s$ quarks in the
proton sea are negatively polarized, whereas the intrinsic $\bar d$
and $\bar s$ antiquarks give zero contributions to the proton spin.
Such a picture is supported by experimental phenomena related to the
proton spin problem: the recent SMC measurement of helicity
distributions for the individual up and down valence quarks and sea
antiquarks, the global fit of different quark helicity contributions
from experimental data, and the negative strange quark helicity from
the $\Lambda$ polarization in $\bar{\nu} {\rm N}$ experiments by the
WA59 Collaboration. The light-cone meson-baryon fluctuation model
also suggests a structured momentum distribution asymmetry for
strange quarks and antiquarks which is related to an outstanding
conflict between two different measures of strange quark sea in the
nucleon. The model predicts an excess of intrinsic $d \bar d$ pairs
over $u \bar u$ pairs, as supported by the Gottfried sum rule
violation. We also predict that the intrinsic charm and anticharm
helicity and momentum distributions are not identical.
 
The intrinsic sea model thus gives a clear picture of quark flavor
and helicity distributions supported qualitatively by a number of
experimental phenomena.  It seems to be an important physical source
for the problems of the Gottfried sum rule violation, the
Ellis-Jaffe sum rule violation, and the conflict between two
different measures of strange quark distributions.

We would like to thank X. Ji and M. Burkardt for helpful conversations.

\end{document}